\documentstyle[12pt,aaspp4]{article}

\lefthead{Bekki \& Freeman}
\righthead{Triaxial halo with figure rotation}

\begin{document}
\title{Extended HI spiral structure and the figure rotation of 
triaxial dark halos}

\author{Kenji Bekki} 
\affil{
School of Physics, University of New South Wales, Sydney 2052, Australia}

\and

\author{Kenneth C. Freeman}
\affil{
Mount Stromlo and Siding Spring Observatories, The Australian National University, 
Private Bag, P.O. Weston Creek, ACT 2611, Australia}

\begin{abstract}
 The HI disk of the blue compact dwarf (BCD) galaxy NGC 2915 extends to
 22 optical scalelengths and shows spiral arms reaching far beyond the
 optical component. None of the previous theories for spiral structure
 provide likely explanations for these very extended spiral arms.  Our
 numerical simulations first demonstrate that such large spiral arms can
 form in an extended  gas disk embedded in a massive triaxial dark
 matter halo with slow figure rotation, through the strong gravitational
 torque of the rotating halo.  We then show that the detailed
 morphological properties of the developed spirals and rings depend
 strongly on the pattern speed of the figure rotation, the shape of the
 triaxial halo, and the inclination of the disk with respect to the
 plane including the triaxial halo's long and middle axes.  These
 results strongly suggest that the dark matter halo of NGC 2915 is
 triaxial and has figure rotation.  Based on these results, we also
 suggest that dynamical effects of triaxial halos with figure rotation
 are important in various aspect of galaxy formation and evolution, such
 as formation of polar ring galaxies, excitation of non-axisymmetric
 structures in low surface-brightness galaxies, and gas fueling to the
 central starburst regions of BCDs.
\end{abstract}


\section{Introduction}

 To understand the  distribution and the nature of dark matter
in galaxies and dynamical effects of dark matter halos
on galaxies has been a longstanding and remarkable problem in galactic astronomy
(e.g., Trimble 1987; Ashman 1992; Salucci \& Persic 1997).
Detailed analysis of rotation curves in variously different galaxies
have played a major role in revealing the radial distributions of dark matter
halos in galaxies (Salucci \& Persic 1997 for a review).
Recently high-resolution rotation curve studies 
for low surface brightness galaxies and dwarfs
have extensively discussed whether or not 
the radial density profiles of dark matter halos predicted from
cold dark matter (CDM) models  (Navarro, Frenk, \& White 1996) 
are consistent with  the observationally inferred profiles
(e.g., de Blok et al. 2001; van den Bosch \& Swaters 2001).

Several attempts have been so far made to reveal the {\it shapes} 
(e.g., the degree of oblateness or triaxiality) of dark matter halos
in galaxies (e.g., Trimble 1987; Ashman 1991).
Following the early attempts to use the kinematics of polar rings for
deriving the three dimensional mass distributions of galaxies 
(Schweizer et al. 1983;  Whitmore et al. 1987), 
Sackett \&  Sparke (1990) tried to give strong constraints
on the shape of a dark matter halo in polar ring galaxy NGC 4650A
by investigating both a rotation curve of a polar ring component
and that of a planner disk one. 
They concluded that the best modeled flattening of the dark halo 
in NGC 4650A is somewhere between E3 and E7. 
Franx, van Gorkom, \& de Zeeuw (1994) analyzed 
both the geometry and the velocity of the  HI gas ring of 
IC 2006 and found a nonsignificant ellipticity of
the gravitational potential ($\sim$ 0.012 $\pm$ 0.026). 

Furthermore, Olling (1995) found that the thickness of the HI gas disk
extending beyond the Holmberg radius in a galaxy
is sensitive to the flattening of the dark matter halo
and proposed that the high resolution HI studies on  the flaring of the outer gas
layers in nearby galaxies enables us to determine the shape of the dark matter halos.
By comparing numerical simulations of dynamical evolution
of the Sagittarius dwarf with observations on the detailed spatial
distribution of the 75 Galactic halo stars,
Ibata et al. (2001) suggested that the Galactic dark halo is
most likely almost spherical in the Galactocentric distance 16 $<$ $R$ $<$ 60 kpc.
Based on the structural  and kinematic properties of the Galactic luminous A-type stars 
revealed by  Hipparcos data (ESA, 1997), 
Cr\'ez\'e et al. (1998) 
argued that there are strong lower limits on the scaleheight 
of any flattened component of the galactic dark halo.

 Although the radial density profiles and the shapes of dark matter
 halos have been extensively discussed by many authors, their {\it
 rotational properties} have been less discussed.  Based on the detailed
 analysis of structure and kinematics of the very extended HI disk
 around NGC 2915, Bureau et al (1999) first suggested that the observed
 spiral-like structures in the HI disk can be formed  by a triaxial halo
 with figure rotation. They furthermore pointed out that the slow
 pattern speed of the figure rotation inferred from the kinematics and
 density distribution of the HI gas is consistent with the pattern
 speeds of rotating triaxial dark halos seen in CDM simulations analysed
 by Pfitzner (2000).  However, it is unclear whether a triaxial dark
 halo with figure rotation is really responsible for the observed
 extended spiral structures in NGC 2915, because of the lack of
 numerical studies of gas dynamics in the gravitational potentials of
 triaxial halos with figure rotation.

 The purpose of this Letter is to demonstrate how non-axisymmetric
 structures (e.g. spirals and bars) can be formed in a gas disk well
 outside the  optical radius of a galaxy embedded in a massive triaxial
 dark matter halo with figure rotation.  We particularly investigate
 how  morphological evolution of outer gas disks depends on the
 structure of the triaxial halos, the  pattern speeds of the figure
 rotation, and the physical properties of the disks (e.g. inclination of
 the disks with respect to the halo and gaseous mass and temperature).

\section{Model}

We consider an extended uniform  gas disk  of a dwarf galaxy
embedded by a triaxial dark matter halo with figure rotation. 
We adopt TREESPH codes described in  Bekki (1997) for hydrodynamical evolution
of galaxies and thereby investigate the dynamical evolution  of the gas disk
under the triaxial dark halo.
All of the following physical parameters are based on observations
on NGC 2915 by Meurer et al (1996) and Bureau et al (1999).
The observed HI extent (the radius within  which the  HI column density 
is observed to be  larger than 5 $\times$ $10^{19}$ cm$^{-2}$) 
is $\sim$ 15 kpc, which is about 5 times
larger than the $B-$band Holmberg radius ($\sim$ 3 kpc) of the host NGC 2915.
This unusually extended HI gas is modeled as an uniform thin gas disk with
the size (represented by $R_{\rm g}$) of 15 kpc and the mass ($M_{\rm g}$) of 10$^8$ $M_{\odot}$.
The gas disk with uniform radial density distribution is represented by
20K SPH particles and each gas particle is 
first placed in the $x$-$y$ plane and given its circular velocity 
(determined by the dark matter halo) at its radius.
The gas disk is then inclined by ${\theta}$ degree  with respect to the $x$-$y$ plane. 
An  isothermal equation of state is used for the gas
with a temperature of $1.2\times 10^3$ K corresponding to a sound speed
of 4 km $\rm s^{-1}$ ($\sim$ 0.075 times the virial velocity of the system).

 This extended and inclined gas disk is assumed to be dynamically affected
 {\it only} by a massive dark matter halo with the mass $M_{\rm DM}$ of
 10$^{10}$ $M_{\odot}$, because the mass of the central stellar components of
 the dwarf (with $M_{\rm B} $ = $-15.9$ mag) is negligibly small compared 
 with that of the dark matter.  We adopt the density distribution of the NFW 
 halo (Navarro, Frenk \& White 1996) suggested from CDM simulations:
 \begin{equation}
 {\rho}(r)=\frac{\rho_{0}}{(r/r_{\rm s})(1+r/r_{\rm s})^2},
 \end{equation} 
 where  $\rho_{0}$ and $r_{\rm s}$ are the central density and the scale
 length of a dark halo, respectively.  It is still controversial whether
 dark matter halos have constant density cores instead of inner steep
 cusps described above (e.g., van den Bosch \& Swaters 2001), so we also
 ran simulations in halos with constant density cores, as in Salucci \&
 Burkert (2000). We find that the resulting structures are essentially
 the same as those for the NFW profiles, so they will not be further
 discussed here.  The scale length $r_{\rm s}$ is treated as a
 parameter, whereas $\rho_{0}$ is chosen such that the total mass of a
 dark halo within the disk radius $R_{\rm g}$ is $M_{\rm DM}$ (=
 10$^{10}$  $M_{\odot}$) for a given $r_{\rm s}$.  We only present the
 results of models with $r_{\rm s}$ = 0.75 kpc, because our results
 depend only weakly on  $r_{\rm s}$.

 We take the isodensity
 surfaces of the dark halo to be triaxial ellipsoids on which the
 Cartesian coordinates ($x$,$y$,$z$) satisfy the following condition
 (Binney \& Tremaine 1987):  \begin{equation} m^2 \equiv
 \frac{x^2}{a^2}+\frac{y^2}{b^2}+\frac{z^2}{c^2} = constant,
 \end{equation} where  $a$, $b$, and $c$ are the parameters which
 determine the two axis ratios of a triaxial body (i.e., long to short
 and long to middle).  In the present study, $a$ is set to be 1 and the
 long-axis is initially coincident with the $x$ axis. Accordingly $b$
 ($\le $ 1) and $c$ ($\le $ 1) are free parameters which determine the
 shapes of triaxial dark matter halos.  The triaxial halo is assumed to
 be rotating as a solid body with the pattern speed of ${\Omega}_{\rm
 p}$.  The figure rotation of triaxial dark matter halos and their
 typical pattern speeds  have been already observed in high-resolution
 collisionless CDM simulations by Pfitzner (2000); see some of the
 results in Bureau et al. (1999).  

By changing the parameters $R_{\rm g}$, $\theta$, $b$, $c$, and ${\Omega}_{\rm p}$,
we investigate morphological evolution of extended gas disks and its dependence
on shapes and rotational properties of triaxial halos.
The parameter values of $R_{\rm g}$, $\theta$, $b$, $c$, and ${\Omega}_{\rm p}$
are 15 kpc, 30$^{\circ}$,  0.8, 0.6,  
and  3.84 km s$^{-1}$ kpc$^{-1}$,
respectively, in  {\it the standard model} (Model 1) which  shows typical behaviors
of gaseous response to triaxial halos and thus described  in detail.
The two  key parameters of spiral arm formation are  
the Toomre's $X$ and $Q$ parameters (Toomre 1964, 1981) and
these are described as
\begin{equation}
X(r)   \equiv \frac{r{\kappa(r)}^2}{2\pi G m \mu(r)},
\end{equation}
where $\kappa(r)$, $m$,  $\mu(r)$ are the epicyclic frequency at a given
radius $r$, the azimuthal wavenumber of the spiral pattern, and the surface
density, respectively, and 
\begin{equation}
Q(r) \equiv \frac{v_{s}(r) \kappa(r)}{\pi G  \mu(r)},
\end{equation}
where $v_{s}(r)$ is the sound speed of the gas.
In this standard model (also in  other  models),
$X$ and $Q$  are rather  large  in the entire disk region
(e.g., 39.3 and 9.4, respectively,  for $m$ = 2 and  $r$ = 15 kpc).
Such gas disks with the $X$ and $Q$ parameters larger than 3 required for
the spiral arm formation (e.g., Binney \& Tremaine 1987) 
are highly unlikely to  form {\rm spontaneously}  spiral arms 
via the ``swing amplification mechanism'' (Toomre 1981).
Therefore, if the spiral arms are formed in our simulations,
this implies that external torque from triaxial halos plays a role in forming
spiral arms in gas disks.
Parameter values for each of 9 models (Model 1 $-$ 9) investigated 
in the present study
are summarized in Table 1. 
There is only one inner Lindblad resonance (ILR) point for all models but Model 4 with
retrograde pattern speed. The points  of ILR ($R_{\rm ILR}$), corotation ($R_{\rm CR}$),
and outer Lindblad resonance ($R_{\rm OLR}$) are given in the table for each model.
In the following, our units of mass, length, and time are
$10^{10}$ M$_{\odot}$ (= M$_{\rm DM}$),
15 kpc, 
and 2.74 $\times$ $10^8$ yr, respectively.

\begin{figure}
\plotone{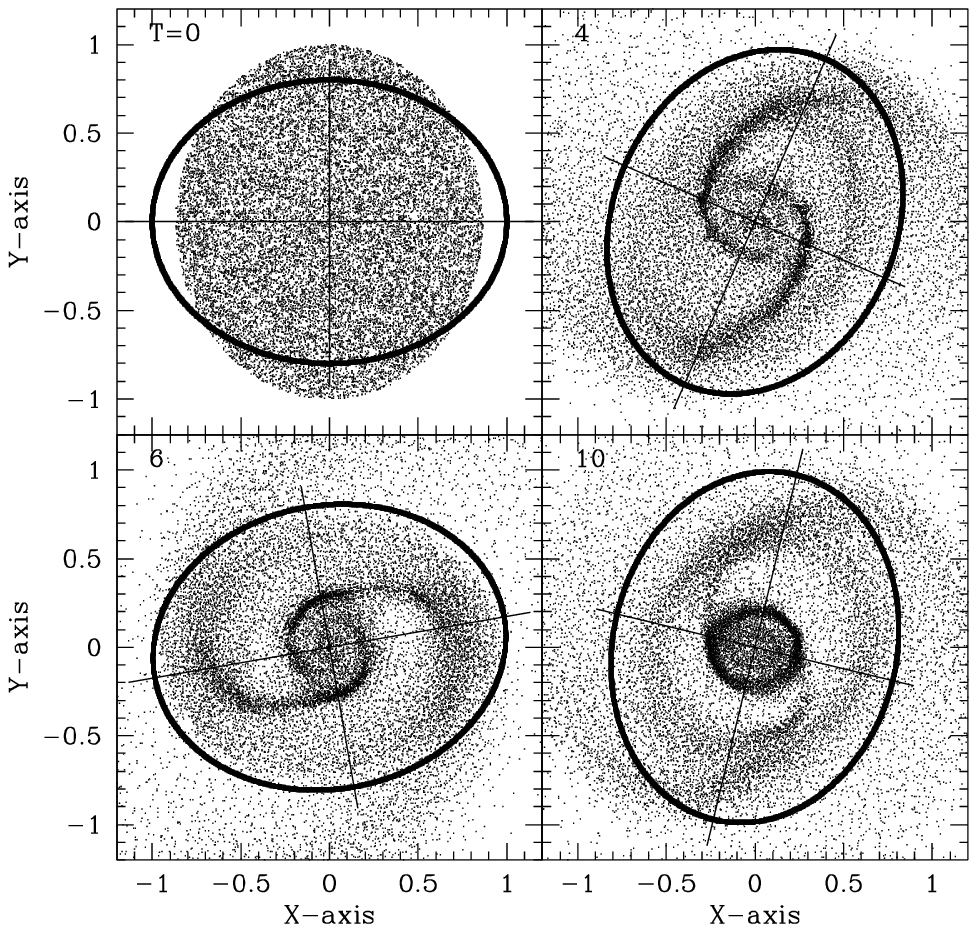}
\figcaption{
Morphological evolution of the gas disk projected onto the $x$-$y$ plane
for the standard model with the disk inclination ($\theta$) of 30$^{\circ}$
with respect to the $x$-$y$ plane. 
The time ($T$) indicated in the upper left
corner of each frame is given in our units (2.74 $\times$ $10^8$ yr) and
each frame (2.4 in our units) measures 36 kpc on a side.
The shape of the triaxial halo projected onto the $x$-$y$ plane (at $R$ = 1.0 in our units,
corresponding to 15 kpc)
is outlined by a thick solid line at each time $T$. 
Both the gas disk and the halo are assumed to rotate counter-clockwise
(The spin axis of the halo is coincident with the $z$ axis).
The long and middle axes  of the halo are represented by thin solid lines.
Note that as the triaxial halo rigidly rotates, two open trailing spiral arms
are formed at $T$ = 4.0 ($\sim$ 1.1 Gyr) owing to  the tidal torque of the halo. 
Note also that these spirals arms finally wind  to form a central gaseous ring
at $T$ = 10  ($\sim$ 2.7 Gyr). Results of models with
the lower inclination angle (0$^{\circ}$ $\le$  $\theta$ $\le$ 30$^{\circ}$)
are essentially the same as that of this model.
\label{fig-1}}
\end{figure}

\placefigure{fig-1}
\placefigure{fig-2}

\section{Results}

Figure 1 describes how gaseous spiral arms are formed, as the triaxial halo
rigidly rotates in the standard model (Model 1).
Owing to the difference in angular speed between the gas disk
and the triaxial halo, the gas disk continuously suffer from  the strong 
tidal force of the halo. 
As a natural result of this, two open trailing arms are 
gradually developed in the entire disk  within 1 Gyr ($T$ = 4). 
These two open arms quickly  wind with each other
to form a central high-density, ring-like
structure at $R$ $\sim$  0.3 (4.5 kpc) in our units ($T$ = 6 and  8).
Annular low-density gaseous regions 
(0.3 $\le$ $R$ $\le$ 0.5 in our units)  
forms just outside the inner ring
because of the disk's angular momentum redistribution caused by  the halo's torque
($T$ = 10).
Thus our simulations first confirms the early suggestion (Bureau et al. 1999) 
that the observed unusually extended spiral arms of NGC 2915
can be due to the tidal torque of the triaxial dark matter halo with figure rotation.

Figure 2 shows that extended remarkable spiral arms can not be formed
in a axisymmetric  dark matter halo with figure rotation (Model 2),
which strongly suggests that  triaxiality is essentially 
important for the formation of gaseous spirals arms in
dark matter halos with figure rotation.
Figure 2 furthermore demonstrates that  even if the self-gravity of
gas is not included in the simulations,
spiral arms are formed in the triaxial dark matter halo (Model 3).
This implies that the physical mechanism of spiral arm formation
in the present model  can be essentially different from 
the ``swing amplification mechanism'' (Toomre 1981) 
which requires self-gravity of disks  
and is a standard model of spiral arm formation.
Sanders \& Huntley (1976) numerically investigated
the dynamical response of  a thin gas disk 
to a rigidly rotating stellar bar
and found that open two-arm trailing spiral structures  can be  formed
owing to the tidal torque of the bar.
They also demonstrated that the orientation of elliptical gas streamlines
formed in the dissipative gas disk
continuously rotates,  and such a pattern of streamlines results in
spiral density enhancement. 
Although their numerical methods and dynamical models are very
different from ours, their proposed mechanism
might well be  essentially the same as that for spiral arm formation
in triaxial dark matter halos with figure rotation.

Detailed morphological properties of gas disks  
are remarkably different between models with different shapes of dark matter halos,
pattern speed of the figure rotation,  sizes  of initial gas disks,
and the inclination angle of disks with respect to the halos.
Figure 2 highlights  important parameter dependences of morphological properties
of gas disks.
Firstly, if the figure rotation is retrograde sense with respect
to the disk's rotation  (Model 4),
the morphological evolution of the  disk is less dramatic,
and consequently remarkable spiral arms can not be seen in the disk
(Instead, ring-like structures can be seen in the central region).
Secondly, very complicated gas rings can be formed in
the model with high inclination angle (Model 5 with $\theta$ = 60$^{\circ}$).
The developed outer parallelogram-like ring and inner small circular ring,
both of which are actually spiral arms in the disk,
are reminiscent of peculiar polar-ring galaxies NGC 660 and NGC 2685. 
Thirdly, more remarkable  spiral arms can be formed in
the model with more bar-like  halo in the $x$-$y$ plane (Model 6).
Fourthly, spiral arm structures are likely to be less remarkable for the
model with slower pattern speed of figure ration (Model 7 and 8),
which confirms  that not only the triaxial shape of the halo
but also the figure rotation
are responsible for spiral arm formation.
Fifthly, it does not depend  strongly on the sizes of initial gas disks
whether or not spiral arms are formed. The model with compact  
gas disk (Model 9 with $R_{\rm g}$ = 0.5)
shows spiral arm structures, which implies
that triaxial halos with figure rotation
can greatly influence even the dynamical evolution of central galaxies.

\placefigure{fig-3}
\placefigure{fig-4}

\section{Discussions and conclusions}

Here are some of the implications of our results.
Firstly, some of the morphological properties of polar ring galaxies
result from gas response to their triaxial dark halo with figure
rotation.  Our models showed that the spatial distributions of gas
disks can be strikingly similar to the double-ring structure observed
in polar ring galaxies ESO 474 - G26 and NGC 2685 (Schweizer et al. 1983)
and to a parallelogram-like structure in  NGC 660 (van Driel et al.
1995).  The simulated structures which appear to be polar ring
components in polar ring galaxies are actually not rings but disks with
spiral arms and with inner low-density gas regions.  This  implies that
{\it some} polar ring components are not rings but `polar disks' with
spiral arms.  NGC 4650A is observed to be such a polar disk with
spiral structure (e.g., Arnaboldi et al.  1997), and accordingly might
be formed by the torque of its rotating triaxial halo.

Secondly, we recall Davies's (1972) suggestion that the HVC complex C
represents an outer extended Galactic spiral arm at relatively high
latitude.  The present simulation successfully produced extended outer
gas arms in a galaxy embedded in a rotating triaxial halo.  We suggest
that high-density regions in the extended gaseous arms excited by the
Galactic triaxial halo could be the formation sites of HVCs.  Extensive
discussions on the origin of HVCs based on the comparison of our
simulations and HIPASS observations will be given in our forthcoming
papers (Bekki \& Freeman 2002).

Thirdly, the morphological properties of low surface brightness (LSB)
disk galaxies can be greatly influenced by triaxial dark matter halos
with  figure rotation, because these LSBs are observed to be dominated
by dark matter (e.g. de Blok \&  McGaugh 1996).  Our simulations show
that even a compact low-mass gas disk can be transformed into spiral
arms by the gravitational torque of its rotating triaxial dark matter
halo.  The observed diversity in non-axisymmetric structures in LSBs
(McGaugh 1992) with possibly large Toomre (1964, 1981) Q and $X$
parameters (i.e. systems in which {\it spontaneous} spiral formation is
highly unlikely) can be due partly to the figure rotation of their
triaxial halos.  Our results also imply that even in high surface
brightness (HSB) galaxies, {\it the outer gas} can be strongly
affected, if the gas is misaligned with the triaxial halos.  We suggest
that the observed  gaseous warps in HSBs can be formed by the halos at
the radius where the strong vertical restoring force of {\it stellar
disks} drops rapidly (e.g., Briggs 1990).

Fourthly, the large-scale torques associated with the rotating triaxial
halos provide a mechanism for fueling the starburst regions of BCDs by
replenishing gas from their outer HI gas reservoirs.


Most dwarf irregular galaxies are observed to have HI gas extending to
$\sim$ twice the Holmberg radius ($R_{\rm H}$). Some of them have gas
out to $4-7 \times R_{\rm H}$ (e.g Hunter 1997).  The morphologies of
this extended HI gas are diverse, ranging from smooth and quiescent
(e.g. NGC 6822; Roberts 1972) to spectacularly complex (e.g. NGC 4449;
Bajaja et al. 1994) and with arms or blobs as in DDO 26 (Hunter \&
Wilcots 2002).  Extended HI with spiral-like structures and arcs is 
seen in spirals like NGC 628 (e.g., Kamphuis \& Briggs 1992).  Our
study strongly suggests that these apparently peculiar structures in
the extended outer HI gas of galaxies can tell us something important
about the shapes and the rotational properties  of dark matter halos.

Systematic comparison between hydrodynamical simulations such as those
presented in this study and high-resolution morphological studies of
unusually extended  HI gas of galaxies might well reveal the dependence
of the pattern speed of rotating triaxial dark halos on the masses,
structures, and environments of galaxies. This could give useful
constraints on theories of galaxies formation.  Finally we would
welcome high-resolution cosmological simulations which provide
better statistics on the incidence of rotating figures of dark matter
halos in low-mass dwarf galaxies.

\acknowledgments

We are  grateful to the anonymous referee for valuable comments,
which contribute to improve the present paper.
K.B.  acknowledges the financial support of the Australian Research Council
throughout the course of this work.
We would like to thank Tim de Zeeuw, Ortwin Gerhard, and Martin Bureau 
for helpful  discussions about rotating triaxial figures.

\clearpage



\begin{figure}
\figcaption{
A collection of morphological properties of eight  different models:
(a) axisymmetric halo model with $b/a$ = 1.0 and $c/a$ = 1.0 (Model 2) 
at $T$ = 10.0 ($\sim$ 2.7 Gyr),
(b) no self-gravity model (Model 3, in which gaseous gravitational interaction is not
included) at $T$ = 4.0 ($\sim$ 1.1 Gyr),
(c) Model 4 at $T$ = 1 (retrograde rotation of the the triaxial halo
with respect to the gas disk rotation),
(d) Model 5 at $T$ = 6 (higher inclination angle of the gas disk, i.e.,
$\theta$ = 60$^{\circ}$,
(e) Model 6 at $T$ = 4 (more strongly barred potential in the $x$-$y$ plane),
(f) Model 7 at $T$ = 4 
(slower angular speed  of ${\Omega}_{\rm P}$ = 0.77 km s$^{-1}$ kpc$^{-1}$), 
(g) Model 8 at $T$ = 4 
(faster angular speed  of ${\Omega}_{\rm P}$ = 7.68 km s$^{-1}$ kpc$^{-1}$),
and (h) Model 9 at $T$ = 4 (compact disk with the size equal to the half
of the disk of the standard model).
Morphological properties  are projected onto the $x$-$y$ plane for Model 2, 3,  4, 6, 7, 8, and 9
and onto the $x$-$z$ one for Model 5.
\label{fig-2}}
\end{figure}

\clearpage

\begin{deluxetable}{cccccccccc}
\footnotesize
\tablecaption{Model parameters 
for gas disks and triaxial dark matter halos \label{tbl-1}}
\tablewidth{0pt}
\tablehead{
\colhead{model} 
& \colhead{${R_{\rm g}}^{a}$} 
& \colhead{$\theta$ ($^{\circ}$)} 
& \colhead{$b/a$} 
& \colhead{$c/a$} 
& \colhead{${{\Omega}_{\rm P}}^{b}$} 
& \colhead{${R_{\rm ILR}}^{a}$} 
& \colhead{${R_{\rm CR}}^{a}$} 
& \colhead{${R_{\rm OLR}}^{a}$} 
& \colhead{comments}}
\startdata
1 & 15 & 30 &  0.8 & 0.6 & 3.84  & 6.8 & 14.2 & 20.7 & the standard model \\
2 & 15 & 30 &  1.0 & 1.0 & 3.84  & 6.8 & 14.2 & 20.7&  axisymmetric \\
3 & 15 & 30 &  0.8 & 0.6 & 3.84  & 6.8 & 14.2 & 20.7& no self-gravity \\
4 & 15 & 30 &  0.8 & 0.6 & -3.84 & - & - & -&  retrograde rotation \\
5 & 15 & 60 &  0.8 & 0.6 & 3.84  & 6.8 & 14.2 & 20.7 &  more inclined disk \\
6 & 15 & 30 &  0.6 & 0.8 & 3.84  & 6.8 & 14.2 & 20.7 &  more bar-like \\
7 & 15 & 30 &  0.8 & 0.6 & 0.77  & 22.5 & 48.1 & 69.8 &  slower figure rotation \\
8 & 15 & 30 &  0.8 & 0.6 & 7.68  & 4.0 & 8.2 & 12.0 &  faster figure rotation \\
9 & 7.5  & 5 &  0.8 & 0.6 & 3.84  &  6.8 & 14.2 & 20.7&  more compact disk\\
\enddata

\tablenotetext{a,b}{in units of kpc and  km s$^{-1}$ kpc$^{-1}$, respectively}
\end{deluxetable}


\begin{thebibliography}{}

\bibitem[]{}
Arnaboldi, M., Oosterloo, T., Combes, F., Freeman, K. C., \& Koribalski, B.
1997, \aj, 113, 585

\bibitem[]{}
Ashman, K. M. 1992, PASP, 104, 1109

\bibitem[]{}
Bajaja, E., Huchtmeier, W. K., \&  Klein, U. 1994, \aap, 285, 385

\bibitem[]{}
Bekki, K. 1997, \apj, 483, 608

\bibitem[]{}
Bekki, K., \& Freeman, K. C. 2002, in preparation

\bibitem[Binney \& Tremaine 1987]{bi87}
Binney, J., \& Tremaine, S., 1987 in Galactic Dynamics.

\bibitem[]{}
Briggs, F. 1990, \apj, 352, 15 

\bibitem[]{}
Bureau, M., Freeman, K. C., Pfitzner, D. W., \&  Meurer, G. R. 1999, \aj, 118, 2158



\bibitem[]{}
Creze, M., Chereul, E., Bienayme, O., \& Pichon, C. 1998, \aap, 329, 920

\bibitem[]{}
Davies, R. D. 1972, \mnras, 160, 381

\bibitem[]{}
de Blok, W. J. G., \&  McGaugh, S. S. 1996, \apj, 469, L89

\bibitem[]{}
de Blok, W. J. G.,  McGaugh, Stacy S., \&  Rubin, V. C. 2001, \aj, 122, 2396


\bibitem[]{}
ESA, 1997, The Hipparcos Catalogue, ESA SP-1200

\bibitem[]{}
Franx, M.,  van Gorkom, J. H., \&  de Zeeuw, T. 1994, \apj, 436, 642




\bibitem[]{}
Hunter, D. A. 1997, PASP, 109, 937

\bibitem[]{}
Hunter, D. A., \& Wilcots, E. M. 2002, prepring (astro-ph/0202149)

\bibitem[]{}
 Ibata, R., Irwin, M, Lewis, G. F., \&  Stolte, A. 2001, \apj, 547, L133

\bibitem[]{}
Kamphuis, J., \&  Briggs, F. 1992, \aap, 253, 335

\bibitem[]{}
McGaugh, S. S. 1992, Ph.D. thesis, Univ. Michigan

\bibitem[]{}
Meurer, G. R., Carignan, C., Beaulieu, S. F., \&  Freeman, K. C.
1996, \aj, 111, 1551

\bibitem[Navarro et al. 1996]{na96}
Navarro, J. F., Frenk, C. S., \& White, S. D. M.
1996, \apj, 462, 563

\bibitem[]{}
Olling, R. P. 1995, \aj, 110, 591

\bibitem[]{}
Pfitzner, D. W. 2000, in preparation

\bibitem[]{}
Roberts, M. S., 1972, in External Galaxies and Quasi-Stellar Object, ed. D. S. Evans
(Dordrecht, Reidel) p.12

\bibitem[]{}
Sackett, P. D., \&  Sparke, L. S. 1990, \apj, 361, 408

\bibitem[]{}
Salucci, P.i, \&  Persic, M., 1997, in Dark and Visible Matter in Galaxies,
ed.  M. Persic and P. Salucci
ASP conf. ser.  Vol. 117, p1,   

\bibitem[]{}
Salucci, P., \&  Burkert, A. 2000, \apj, 537, L9

\bibitem[]{}
Sanders, R. H. \&  Huntley, J. M. 1976, \apj, 209, 53

\bibitem[]{}
Schweizer, F.,  Whitmore, B. C., \&  Rubin, V. C. 1983, \aj, 88, 909

\bibitem[]{}
Toomre, A. 1964, \apj, 139, 1217

\bibitem[]{}
Toomre, A. 1981, in  The structure and evolution of normal galaxies, ed. S. M. Fall
\& D. Lynden-Bell (Cambridge University Press) p111

\bibitem[]{}
Trimble, V. 1987, ARAA, 25, 425

\bibitem[]{}
van den Bosch, F. C., \&  Swaters, R. A. 2001, \mnras, 325, 1017

\bibitem[]{}
van Driel, W. et al. 1995, \aj, 109, 942


\bibitem[]{}
Whitmore, B. C., McElroy, D. B., \&  Schweizer, F. 1987, \apj, 314, 439























\end{thebibliography}
\end{document}